\documentstyle[prl,aps]{revtex}
\input BoxedEPS.tex
\SetOzTeXEPSFSpecial
\HideDisplacementBoxes

\begin{document}

%\draft{}
\twocolumn[\hsize\textwidth\columnwidth\hsize
          \csname @twocolumnfalse\endcsname
\title{Isotope effects and possible pairing mechanism in optimally doped 
cuprate superconductors}
\author{Guo-meng Zhao$^{1,2}$, Vidula Kirtikar$^{2}$, and Donald E. 
Morris$^{2}$} 
\address{$^{1}$Physik-Institut der Universit\"at Z\"urich,
CH-8057 Z\"urich, Switzerland~\\
$^{2}$Morris Research, Inc., 44 Marguerito Road, Berkeley, CA 96707, 
USA}

\maketitle
\widetext

\begin{abstract}

We have studied the oxygen-isotope effects on $T_{c}$ and in-plane penetration 
depth $\lambda_{ab}(0)$ in an optimally doped 3-layer cuprate 
Bi$_{1.6}$Pb$_{0.4}$Sr$_{2}$Ca$_{2}$Cu$_{3}$O$_{10+y}$ 
($T_{c} \sim$ 107 K). 
We find a small oxygen-isotope effect on $T_{c}$ ($\alpha_{O}$  = 0.019), and 
a substantial effect on $\lambda_{ab} (0)$ ($\Delta \lambda_{ab} 
(0)/\lambda_{ab} (0)$ =  
2.5$\pm$0.5$\%$). The present results along with the previously 
observed isotope effects in single-layer and double-layer cuprates indicate that 
the isotope exponent $\alpha_{O}$ in 
optimally doped cuprates is small while the isotope effect on the 
in-plane effective supercarrier mass 
is substantial and nearly independent of the number of the CuO$_{2}$ 
layers. A plausible pairing mechanism 
is proposed to explain the isotope effects, high-T$_{c}$ 
superconductivity and tunneling spectra in a consistent way.

\end{abstract}
\vspace{1cm}
\narrowtext
%\newpage
]
The pairing mechanism responsible for high-$T_{c}$ superconductivity is 
still controversial. In conventional superconductors, a strong effect of 
changing ion mass $M$ on the transition temperature $T_{c}$ implies 
that lattice vibrations (phonons) play an important role in the 
microscopic mechanism of superconductivity. An isotope exponent 
$\alpha$ (= $- d\ln T_{c}/d\ln M$) of about 0.5 is consistent with the 
phonon-mediated BCS theory. A nearly zero oxygen-isotope effect 
($\alpha_{O}$ $\simeq$ 0.03) was earlier observed in a double-layer cuprate 
superconductor YBa$_{2}$Cu$_{3}$O$_{7-y}$ which is optimally doped 
($T_{c} \sim$ 92 K) \cite{Batlogg,Don88}. Such a small isotope effect 
might suggest that phonons should not be important to the pairing 
mechanism. On the other hand, large 
oxygen-isotope shifts were later observed in several underdoped 
cuprate superconductors 
\cite{Crawford90,Crawford2,Bornemann92,Franck93,ZhaoLSCO,ZhaoNature97,ZhaoJPCM}. 
Further, three indirect experiments have 
consistently demonstrated that the difference in the hole densities of the $^{16}$O and $^{18}$O samples 
is smaller than 0.0002 per Cu site \cite{ZhaoLSCO,ZhaoNature97,ZhaoJPCM}. 
Moreover, a quantitative data analysis on the isotope-exchanged 
YBa$_{2}$Cu$_{3}$O$_{6.94}$ \cite{ZhaoYBCO} suggested that there is 
a negligible oxygen-isotope effect on the supercarrier density $n_{s}$. 

Since muon-spin rotation experiments \cite{Uemura89} showed that 
$T_{c}$ is approximately proportional to $n_{s}/m^{**}_{ab}$ in 
deeply underdoped cuprates (where $m^{**}_{ab}$ is the in-plane 
effective supercarrier mass), a large oxygen-isotope shift of $T_{c}$ 
observed in this doping regime should arise from a large oxygen-isotope 
effect on $m^{**}_{ab}$. Indeed, several independent experiments 
\cite{ZhaoLSCO,ZhaoNature97,ZhaoJPCM,HoferPRL} have consistently 
demonstrated that both the average supercarrier mass $m^{**}$ and the 
in-plane supercarrier mass $m^{**}_{ab}$ strongly depend on the oxygen 
isotope mass in underdoped cuprates. Such an unconventional isotope 
effect suggests that there exist polaronic charge carriers, which are 
condensed into supercarriers in the superconducting state. This 
appears to give a support to a theory of (bi)polaronic 
superconductivity \cite{ale}. On the other hand, within this theory, it 
is difficult to explain a small isotope shift of $T_{c}$ and a large 
reduced energy gap (i.e., 
2$\Delta (0)/k_{B}T_{c}$ $>$ 6) observed in optimally doped cuprates where 
single-particle excitation gap vanishes above $T_{c}$ \cite{Miyakawa}. 
Therefore, an alternative theoretical approach is required to explain 
superconductivity in optimally doped and overdoped cuprates. A 
possibly correct pairing mechanism should be able to consistently explain a small 
isotope shift of $T_{c}$, a substantial isotope effect on 
the supercarrier mass \cite{ZhaoNature97,ZhaoYBCO}, and a large reduced 
energy gap. 

Here we report the observation of the oxygen-isotope effects on $T_{c}$ 
and in-plane penetration 
depth $\lambda_{ab} (0)$ in a 3-layer cuprate 
Bi$_{1.6}$Pb$_{0.4}$Sr$_{2}$Ca$_{2}$Cu$_{3}$O$_{10+y}$ ($T_{c} \sim$ 107 K). 
We find a small oxygen-isotope effect on $T_{c}$ and a 
substantial effect on $\lambda_{ab} (0)$. We propose a possible 
theoretical model which is able to consistently explain these 
isotope effects and the large reduced energy gap.

Samples of Bi$_{1.6}$Pb$_{0.4}$Sr$_{2}$Ca$_{2}$Cu$_{3}$O$_{10+y}$ 
were prepared from high purity Bi$_{2}$O$_{3}$, PbO, SrCO$_{3}$, 
CaCO$_{3}$, and CuO. The 
samples were ground, pelletized and fired at 865/855/845 $^{\circ}$C for 
37/40/44 h in air with two intermediate grindings. Two samples were prepared 
under nearly the same heat treatment. To ensure that the samples have small 
grain size and enough porosity, they were reground thoroughly, 
pelletized and annealed in flowing oxygen at 600 $^{\circ}$C for 10 h.

Two pelletized samples were broken into halves (producing two 
sample pairs), and the halves were subject to the $^{16}$O and $^{18}$O isotope 
diffusion, which was conducted in two parallel quartz tubes separated 
by about 2 cm \cite{Bornemann92,ZhaoYBCO,ZhaoLSCO}.  The diffusion was carried out for 68 
h at 600 $^{\circ}$C and oxygen pressure of 0.8 bar for sample pair I, and 
for 40 h at 650 $^{\circ}$C and oxygen pressure of 1 bar for sample pair II. 
The cooling rate was 30 $^{\circ}$C/h. The oxygen isotope enrichments 
were determined from the weight changes of both $^{16}$O and $^{18}$O 
samples. The $^{18}$O samples had 85$\pm$5$\%$ $^{18}$O and 15$\pm$5$\%$ $^{16}$O.

The susceptibility was measured with a Quantum Design SQUID 
magnetometer. The field-cooled, measured-on-warming susceptibility was measured in a field of 15 Oe for sample pair I, 
and 10 Oe for sample pair II. The temperature measurements 
were performed with a platinum resistance thermometer (Lakeshore 
PT-111) placed in direct contact with the sample and driven  by 
microprocessor controlled ac bridge in the SQUID magnetometer. The 
resolution is 2.5 mK and reproducibility is 10 mK at 77 K after 
cycling to room temperature \cite{Bornemann92,ZhaoYBCO}.
\begin{figure}[htb]
%\vspace{2cm}
    \ForceWidth{7cm}
	\centerline{\BoxedEPSF{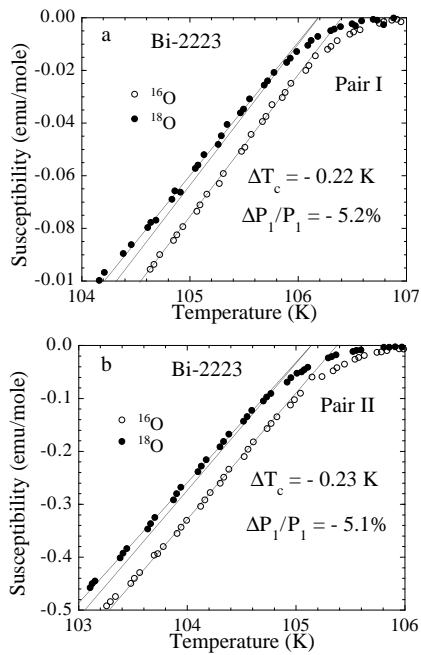}}
	\vspace{0.3cm}
	\caption[~]{The susceptibility near $T_{c}$ for the $^{16}$O and $^{18}$O 
samples of Bi$_{1.6}$Pb$_{0.4}$Sr$_{2}$Ca$_{2}$Cu$_{3}$O$_{10+y}$ 
(Bi-2223): (a) pair I; (b) pair II.}
	\protect\label{Fig.1}
\end{figure} 
In Fig.~1, we show the susceptibility near $T_{c}$ for the $^{16}$O 
and $^{18}$O samples of 
Bi$_{1.6}$Pb$_{0.4}$Sr$_{2}$Ca$_{2}$Cu$_{3}$O$_{10+y}$: (a) pair I; (b) pair II.  In all 
the cases, the $T_{c}$ for the $^{18}$O samples is  0.22$\pm$0.01 K lower than 
for the $^{16}$O samples. Extrapolating to 100$\%$ exchange, we calculate the 
isotope exponent $\alpha_{O} = - d\ln T_{c}/d\ln M_{O}$ = 0.019$\pm$001. 
We also note that there is a well-defined linear portion on the 
transition curve $\sim$ 1 K below the diamagnetic onset temperature. It is evident that 
the slope of the linear portion (denoted by $P_{1}$) for the  $^{18}$O samples 
is 5.0$\pm$1.0$\%$ smaller than for the $^{16}$O samples, that is, 
$\Delta P_{1}/P_{1} = - 5.0\pm1.0\%$.

For comparison, the results for single-layer 
La$_{1.85}$Sr$_{0.15}$CuO$_{4}$ (LSCO) and double-layer 
YBa$_{2}$Cu$_{3}$O$_{6.94}$ (YBCO) compounds are reproduced in Fig.~2. It is 
clear that $\Delta P_{1}/P_{1} = - 5.5\pm1.0\%$ for LSCO and $- 6.8\pm1.0\%$
for YBCO. Comparing Fig.~2 with Fig.~1, one can see that the isotope effect 
on $P_{1}$ is nearly the same for all three compounds within the 
experimental uncertainty. It is 
also interesting to note that 
the isotope effect on $T_{c}$ decreases monotonically with increasing $T_{c}$.

The observed oxygen-isotope effect on the slope $P_{1}$ is 
caused by the dependence of the penetration depth on 
the oxygen mass. For nearly isotropic materials with 
$\lambda_{ab}(0) \sim \lambda_{c}(0) \sim \lambda (0)$, it was shown 
that \cite{ZhaoYBCO}:
$\Delta P_{1}/P_{1} = - \Delta T_{c}/T_{c} - 2\Delta\lambda (0)/\lambda 
(0)$.  For highly anisotropic materials such as cuprates, a relation $\lambda_{c}(T) >> 
\lambda_{ab}(T) > R$ (where $R$ is the maximum particle size) holds near 
$T_{c}$ so that the diamagnetic signal is proportional to 
$1/\lambda^{2}_{ab}(T)$ \cite{Buzdin}. Then one readily finds that 
\begin{equation}
\Delta P_{1}/P_{1} \simeq - \Delta T_{c}/T_{c} - 2\Delta\lambda_{ab} 
(0)/\lambda_{ab} (0).
\end{equation}
From Fig.~1 and Fig.~2, we obtain $\Delta\lambda_{ab} 
(0)/\lambda_{ab} (0)$ = 3.2$\pm$0.7$\%$ using Eq.~1. This is in remarkably good 
agreement with the recent muon spin rotation experiments on 
the oxygen-isotope exchanged YBa$_{2}$Cu$_{3}$O$_{6.96}$, which show 
that $\Delta\lambda_{ab} 
(0)/\lambda_{ab} (0)$ = 2.5$\pm$0.5$\%$ \cite{Kellerprep}. 
\begin{figure}[htb]
%\vspace{2cm}
    \ForceWidth{7cm}
	\centerline{\BoxedEPSF{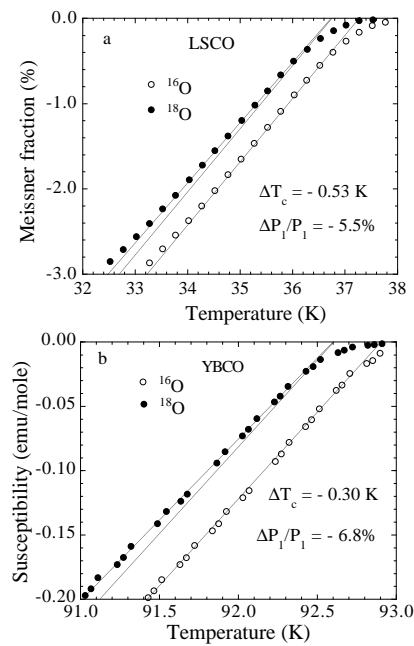}}
	\vspace{0.3cm}
	\caption[~]{The susceptibility data near $T_{c}$ for the $^{16}$O and $^{18}$O 
samples of (a) La$_{1.85}$Sr$_{0.15}$CuO$_{4}$ (LSCO) and (b) 
YBa$_{2}$Cu$_{3}$O$_{6.94}$ (YBCO). After 
Refs.\cite{ZhaoNature97,ZhaoYBCO}. }
	\protect\label{Fig.2}
\end{figure}

Since both $n$ and $n_{s}$ are independent of the isotope mass as 
discussed above, the observed oxygen-isotope effect on the in-plane penetration depth is 
caused by the isotope dependence of $m^{**}_{ab}$.  The 
substantial isotope effect on $m^{**}_{ab}$ may suggest that charge 
carriers in the optimally doped cuprates remain polaronic nature, and that 
those polaronic carriers are condensed into supercarriers in the superconducting state.

Now a question arises: why is the isotope effect on $m^{**}_{ab}$ 
substantial while the isotope shift of $T_{c}$ is very small in these 
optimally doped cuprates? In order to answer this question, we should 
first find out which phonon modes are strongly coupled to doped holes in 
these materials. Inelastic neutron scattering experiments 
\cite{McQueeney,Petrov} show that the 
Cu-O bond stretching mode in the CuO$_{2}$ planes are strongly 
coupled to the doped holes. The average frequency of this mode is 
about 75 meV in La$_{1.85}$Sr$_{0.15}$CuO$_{4}$ \cite{McQueeney}, and about 60 meV in 
YBa$_{2}$Cu$_{3}$O$_{6.92}$ \cite{Petrov}. Such a strong electron-phonon 
coupling should be also manifested in tunneling spectra, as it is the 
case in conventional superconductors. Although it is difficult to obtain reliable tunneling spectra for 
cuprates due to a short coherent length, there are two high-quality tunneling spectra for slightly overdoped YBCO 
\cite{Wei} and Bi$_{2}$Sr$_{2}$CaCu$_{2}$O$_{8+y}$ (BSCCO) \cite{Gonnelli}.
\begin{figure}[htb]
%\vspace{2cm}
    \ForceWidth{7cm}
	\centerline{\BoxedEPSF{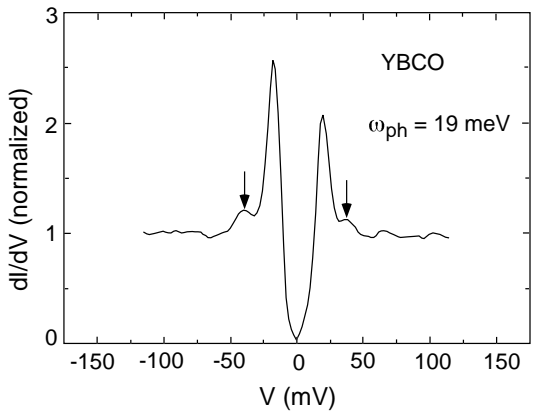}}
	\vspace{1cm}
	\caption[~]{The normalized conductance data for a scanning tunneling 
microscopy (STM) tunnel junction on a slightly overdoped 
YBa$_{2}$Cu$_{3}$O$_{7-y}$ (YBCO) crystal 
with a Pt-Ir tip at 4.2 K. After \cite{Wei}. }
	\protect\label{Fig3}
\end{figure}
In Fig.~3, we show normalized conductance data for a scanning tunneling 
microscopy (STM) on a slightly overdoped YBCO crystal 
with a Pt-Ir tip at 4.2 K \cite{Wei}. The crystal has $T_{c}$ $\simeq$ 90 K with 
$\sim$ 1 K transition width \cite{Wei}. Since the presence of oxygen 
vacancies in the CuO chains can lead to residual 
density of states and to zero-bias conductance in the superconducting 
state, a negligible zero-bias conductance in the spectrum suggests that 
the spectrum represents the intrinsic density of states contributed only from 
the CuO$_{2}$ planes. It is striking that the strong coupling 
features similar to that in conventional superconductors can be clearly 
seen in the spectrum (as indicated by the arrows). The 
strong-coupling features correspond to a strong coupling between 
charge carriers and the phonon mode with $\omega_{ph}$ = 19 meV. 
The phonon density of states in an optimally doped YBCO also reveals a 
large peak at about 20 meV \cite{Renker}.

In Fig.~4, we plot the electron-phonon spectral density 
$\alpha^{2} F(\omega)$ for an optimally-doped 
BSCCO crystal, which was 
extracted from an SIS break-junction spectrum \cite{Gonnelli}. A strong 
coupling feature at an 
energy of about 20 meV is clearly seen. This feature also corresponds to 
the large peak in the phonon density of states at about 20 meV (see 
open circles). In 
addition to a strong coupling feature at 20 meV, there is another 
strong coupling feature at about 73 meV, which corresponds to the 
phonon energy of the Cu-O bond stretching mode discussed above. Therefore, these 
tunneling spectra (Fig.~3 and 4) consistently suggest that both the 
low-energy phonon mode at about 20 meV and the high-energy phonon mode 
at about 73 meV are strongly coupled to conduction electrons in these 
double-layer compounds.

\begin{figure}[htb]
%\vspace{2cm}
    \ForceWidth{7cm}
	\centerline{\BoxedEPSF{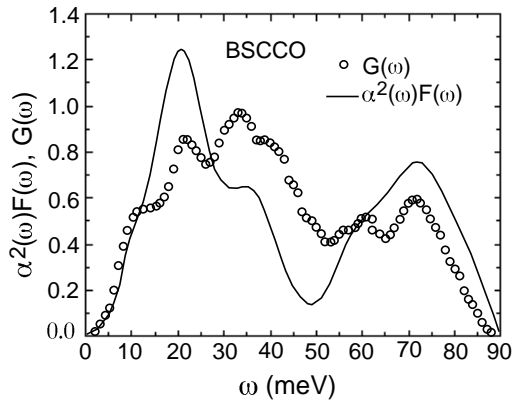}}
	%\vspace{1cm}
	\caption[~]{The electron-phonon spectral density 
$\alpha^{2} F(\omega)$ for an optimally doped 
Bi$_{2}$Sr$_{2}$CaCu$_{2}$O$_{8+y}$ (BSCCO) crystal, which was deduced 
from an SIS break-junction spectrum \cite{Gonnelli}. }
	\protect\label{Fig4}
\end{figure}
A theoretical approach to a strong electron-phonon coupling system 
depends not only on the
adiabatic ratio $\omega/E_{F}$ but also on the coupling strengths with
different phonon modes.
When the Fermi energy $E_{F}$ is smaller than these strong-coupling
phonon energies, the phonon-induced effective 
interaction between carriers is nonretarded so that the real-space 
pairing (e.g., intersite bipolaron formation) becomes possible 
\cite{ale,tru}. This 
should be the case for the doping level $x$ $\leq$ 0.10 in 
La$_{1-x}$Sr$_{x}$CuO$_{4}$ \cite{ZhaoJPCM}. In contrast, $E_{F}$ in 
doped oxygen-hole bands may 
lie in between 
20 and 73 meV in the optimally doped and overdoped regimes, so that the pairing 
interaction becomes retarded for the low-energy phonons, and remains 
nonretarded for the high-energy phonons. The retarded electron-phonon 
coupling for the low-energy phonons could be treated within the Migdal 
approximation, while the nonretarded electron-phonon coupling for the 
high-energy phonons should be modeled separately within the polaron 
theory. This theoretical approach has been successfully applied to 
fullerenes \cite{Alex96}. 
The strong coupling 
between doped holes and the high-energy phonons leads to a polaronic 
mass enhancement and to an attractive nonretarded potential between doped holes. 
Effectively, the polaronic holes could then form k-space Cooper pairs 
by interacting with the low-energy phonons. The problem could 
thus be solved within 
Eliashberg equations with an effective electron-phonon spectral density 
for the low-energy phonons and a negative Coulomb pseudopotential 
produced by the high-energy phonons and other high-energy bosonic 
excitations of purely electronic origin (e.g., spin fluctuations, excitons, 
and plasmons). Within this simplified approach, the effective 
electron-phonon coupling 
constant $\lambda_{ep}$ for the low-energy phonons is enhanced by 
a factor of $f_{p} = \exp (g^{2})$. Here $g^{2} = A/\omega_{H}$, $A$ 
is a constant, and $\omega_{H}$ is the frequency of the high-energy 
phonon mode.  The effective Coulomb 
pseudopotential $\mu^{*}$ is negative and also proportional to $f_{p}$.

For slightly overdoped BSCCO, $g^{2}$ can be evaluated from the 
mid-infrared optical conductivity which exhibits a maximum at 
$E_{m}$ $\simeq$ 0.12 eV \cite{Quijada}. With $E_{m}$ = 0.12 eV, 
$\hbar\omega_{H}$ = 
75 meV,  we find $g^{2} = E_{m}/(2\hbar\omega_{H})$ = 0.8, leading to 
$f_{p}$ = 2.2. From the spectral density shown in Fig.~4, we can 
extract the effective electron-phonon coupling constant  
$\lambda_{ep}$ for the low-energy phonon mode, that is, 
$\lambda_{ep}$ $\simeq$ 2.6. If there were no polaronic mass 
enhancement due to the high-energy phonons, the coupling constant 
contributed from the low-energy phonons would be about 
1.2. With $\mu^{*}$ = 0.1 and $\lambda_{ep}$ = 1.2, we calculate 
$T_{c}$ = 18 K according to a $T_{c}$ formula
\begin{equation}\label{Te6}
k_{B}T_{c} = 0.25\hbar\sqrt{<\omega^{2}>}[\exp (2/\lambda_{eff}) - 1]^{-1/2},
\end{equation}
where
\begin{equation}\label{Te7}
\lambda_{eff} = (\lambda_{ep}-\mu^{*})/[1 + 2\mu^{*} + 
\lambda_{ep}\mu^{*}t(\lambda_{ep})],
\end{equation}
The function $t(\lambda_{ep})$ is plotted in Fig.~2 of 
Ref.\cite{Kresin}. In the present case, $\hbar\sqrt{<\omega^{2}>}$ is 
contributed only from the low-energy phonons and equal to 20 meV. Therefore, without the 
high-energy phonons, $T_{c}$ would not be higher than 20 K. The 
high-energy phonons not only enhance $\lambda_{ep}$ by a factor of 
2.2, but also reduce $\mu^{*}$ substantially \cite{Alex96}. It has recently been shown 
that $\mu^{*}$ in cuprates becomes 
negative (i.e., $\mu^{*} \simeq -$0.05) due to the presence of low-energy electronic collective 
modes (acoustic plasmons) in layered conductors \cite{Bill}. Since the high-energy phonon 
mode reduces $\mu^{*}$ further, it is likely that the value of 
$\mu^{*}$ should be in the range of $-$0.1 to $-$0.2. If we take $\mu^{*} = 
-$0.15, $\lambda_{ep} = 2.6$ (see above), we can get $T_{c}$ = 95 K. This leads 
to $k_{B}T_{c}/(\hbar\sqrt{<\omega^{2}>})$ = 0.41, and $2\Delta 
(0)/k_{B}T_{c}$ $\simeq$ 7 according to the known
relation between $k_{B}T_{c}/(\hbar\sqrt{<\omega^{2}>})$ and $2\Delta 
(0)/k_{B}T_{c}$ \cite{CarbotteRev}. The calculated reduced energy gap 
is in good agreement with experiment. We would like to mention that 
the present calculation is valid for an isotropic s-wave gap, which 
should be the case for the polaronic oxygen-hole bands near (0,$\pm\pi$) and 
($\pm\pi$, 0) regions. In addition to the polaronic oxygen holes, there are 
electron-like free carriers which could condense into supercarriers through 
interband scattering. The pairing symmetry of the two-carrier system 
may be an extended s-wave with eight line nodes \cite{ZhaoPRB}. 

Now we can calculate the total isotope exponent $\alpha$ 
using Eqs.~\ref{Te6} and \ref{Te7}, and the relations: $\lambda_{ep} \propto f_{p}$, $\mu^{*} 
\propto f_{p}$, $t(\lambda_{ep})\simeq 1.8/\lambda_{ep}$. The calculated  
total isotope exponent is $\alpha \simeq$ 0. The nearly zero isotope exponent is due to the 
fact that the isotope dependencies of $\lambda_{ep}$ and $\mu^{*}$ 
(arising from the polaronic effect) cancel out the isotope effect on the 
prefactor of Eq.~\ref{Te6}. Moreover, the formation of polaronic Cooper 
pairs naturally explains the sizable isotope effect on the 
supercarrier mass.

In summary, we report the observation of the oxygen-isotope effects 
in the 3-layer cuprate 
Bi$_{1.6}$Pb$_{0.4}$Sr$_{2}$Ca$_{2}$Cu$_{3}$O$_{10+y}$. The present results 
along with the previously 
observed isotope effects in single-layer and double-layer cuprate 
superconductors indicate that the isotope effect on $T_{c}$ in 
optimally doped cuprates is small while the isotope effect on the effective supercarrier mass 
is substantial. These isotope effects and tunneling spectra observed in optimally 
doped cuprates can be consistently explained within a scenario 
where polaronic oxygen holes are bound into Cooper pairs.

\end{document}